\documentclass[%
 reprint,
superscriptaddress,
showkeys,
nofootinbib,
 amsmath,amssymb,
 aps,
floatfix,
]{revtex4-2}
\usepackage{accents}
\usepackage{graphicx}
\usepackage{dcolumn}
\usepackage{bm}
\usepackage[utf8]{inputenc}
\usepackage{tabularx}
\usepackage[dvipsnames]{xcolor}
\usepackage{units}
\usepackage[euler]{textgreek}
\usepackage{upgreek}
\usepackage{derivative}
\usepackage{amsmath}
\usepackage{physics}
\usepackage{amssymb}
\usepackage[colorlinks=true,linkcolor=blue!50!black,urlcolor=blue!50!black,citecolor=blue!50!black]{hyperref}

\graphicspath{ {img/} }

\begin{document}

\preprint{APS}

\title{Dark Matter-Electron Scattering Search Using Cryogenic Light Detectors}

\author{V.~Zema}
\email{vanessa.zema@mpp.mpg.de}
\affiliation{Max-Planck-Institut f\"ur Physik, 85748 Garching - Germany}

\author{P.~Figueroa}
\email{pablof@mpp.mpg.de}
\affiliation{Max-Planck-Institut f\"ur Physik, 85748 Garching - Germany}

\author{G.~Angloher}
\affiliation{Max-Planck-Institut f\"ur Physik, 85748 Garching - Germany}

\author{M.~R.~Bharadwaj}
\affiliation{Max-Planck-Institut f\"ur Physik, 85748 Garching - Germany}

\author{T.~Frank}
\affiliation{Max-Planck-Institut f\"ur Physik, 85748 Garching - Germany}

\author{M. N. Hughes~}
\affiliation{Max-Planck-Institut f\"ur Physik, 85748 Garching - Germany}

\author{M.~Kellermann}
\affiliation{Max-Planck-Institut f\"ur Physik, 85748 Garching - Germany}

\author{F.~Pr\"obst}
\affiliation{Max-Planck-Institut f\"ur Physik, 85748 Garching - Germany}

\author{K.~Sch\"affner}
\affiliation{Max-Planck-Institut f\"ur Physik, 85748 Garching - Germany}

\author{K.~Shera}
\affiliation{Max-Planck-Institut f\"ur Physik, 85748 Garching - Germany}

\author{M.~Stahlberg}
\affiliation{Max-Planck-Institut f\"ur Physik, 85748 Garching - Germany}

\begin{abstract}
The CSC (cryogenic scintillating calorimeter) technology devoted to rare event searches is reaching the sensitivity level required for the hunt of dark matter-electron scatterings. Dark matter-electron interactions in scintillating targets are expected to stimulate the emission of single photons, each of energy equal to the target electronic band gap. The electronic band gap in scintillators like NaI/GaAs is of $\mathcal{O}$(eV). The search for this signal can be done by an array of cryogenic light detectors with eV/sub-eV energy resolution. In this work, we describe the detection principle, the detector response and the envisioned detector design to search for dark matter interacting with electrons via the measurement of the scintillation light at millikelvin. 
First sensitivity projections are provided, which show the potential of this research.
\end{abstract}

\maketitle



\section{Introduction}
The consistency between the cosmological data and the $\Lambda$CDM model~\cite{aghanim2020planck, Workman:2022ynfbbn, Peebles:2022akh} strongly points towards the existence of a non-baryonic matter component of our Universe, which requires an extension of the Standard Model of particles~\cite{Workman:2022ynf}.
Several theories which predict the interaction of DM with ordinary matter have been proposed (see e.g.~\cite{Workman:2022ynfdm} for a recent review). Among these, there are models which have successfully predicted the correct relic abundance, with light DM candidates below the GeV scale and a new gauge boson/mediator~\cite{boehm2004scalar, Feng:2008ya, Hooper:2008im, Lin:2011gj, hochberg2014mechanism, Kuflik:2015isi, PhysRevD.99.115009, boehm2021scalar,Zurek:2024qfm}. For example, the existence of an additional vector particle, a \textit {dark photon}, acting as mediator to a new particle portal, has been postulated~\cite{holdom1986two} and extensively investigated as possible solution to the DM problem~\cite{An:2014twa, Mitridate:2021ctr, Essig:2013lka, Alexander:2016aln, Battaglieri:2017aum}. At the energy scale probed by DM direct detection experiments, the dark photon can generate an effective coupling of DM with electrons, which implies that DM-electron interactions could be observed in such experiments~\cite{An:2014twa, Essig_2012, Bloch:2016sjj, Blanco:2022cel, Essig:2022dfa}. Models beyond the dark-photon can also produce the same phenomenology~\cite{Catena:2022fnk}.~Diverse experimental observations from cosmology, astrophysics and earth-bounded DM direct detection experiments have been used to constrain the coupling and mass parameter space of DM interacting with electrons~\cite{Knapen:2017xzo, Chang:2018rso, Buen-Abad:2021mvc, Chu:2023jyb}. DAMIC-M~\cite{DAMIC-M:2023gxo}, DarkSide-50~\cite{DarkSide:2022knj}, EDELWEISS~\cite{PhysRevLett.125.141301}, PANDA-X~\cite{PandaX-II:2021nsg}, SENSEI~\cite{SENSEI:2020dpa}, SuperCDMS~\cite{PhysRevLett.121.051301} and XENON-1T~\cite{XENON:2021qze} reached the sensitivity to DM-electron scatterings and are probing this region, providing new constraints. While the sensitivity to DM-nuclei interactions achieved by DM direct detection experiments is such that the neutrino fog is now at the door~\cite{akerib2022snowmass2021, PhysRevLett.126.091301}, a large parameter space corresponding to DM-electron interactions is still to be disclosed~\cite{Carew:2023qrj}.\\
In this landscape, the contribution of the experiments based on the cryogenic scintillating calorimeters (CSCs) is missing. {\color{black} CSCs with transition edge sensors have been developed by CRESST (in phase II and III~\cite{CRESST-II:2014ezs, CRESST:2019jnq,CRESST:2024cpr}) for the search of DM-nuclei interactions} and in the recent years also by COSINUS~\cite{COSINUS:2023kqd} for the search of DM-nuclei interactions in NaI targets and by NUCLEUS~\cite{NUCLEUS:2019igx} for the search of neutrino-nuclei coherent scatterings. The employment of the CSCs in the search for DM-electron interactions, which was first discussed in~\cite{Derenzo:2016fse}, requires a dedicated optimized detector design and theoretical framework. A similar effort is underway by the TESSERACT project using GaAs absorbers~\cite{Essig:2022dfa}. In this study, we present preliminary investigations of the detector response and expected signature of the signal in CSCs, as well as on the sensitivity of CSCs to DM-electron interactions in presence of a known monotonically decaying background.

\section{DM-electron interaction}
\label{sec:DMe}
The theory of DM-electron interactions in direct detection experiments has been extensively studied in the recent years in numerous works~\cite{Essig_2012, Essig:2015cda, Lee:2015qva, Derenzo:2016fse, Catena:2019gfa, Knapen:2021bwg, Catena:2021qsr, Hochberg:2021pkt}. Without delving into the full treatment here, we highlight the most relevant aspects for this work. \\

\begin{figure}[ht]
\includegraphics[width=\linewidth]{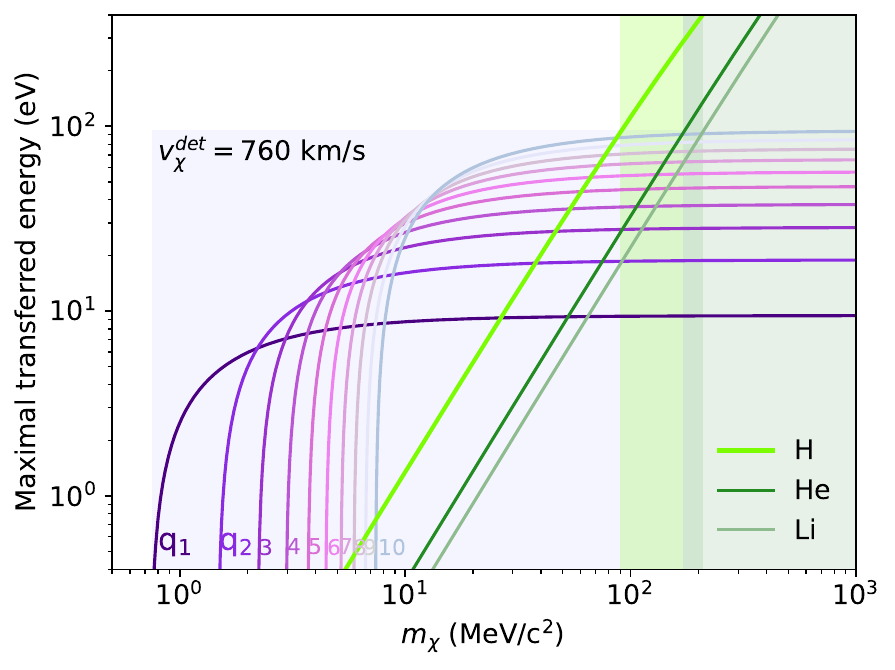}
\caption{\label{fig:kinematics} The maximal recoil energy in DM-electron scattering as a function of the DM mass for different transferred momenta, $q_i, i=1,\dots,10$, depicted using violet ($q_1$) to light blue ($q_{10}$) curves. The maximal nuclear recoil energy for DM-nuclei elastic scattering for three light targets, H, He and $^6$Li, depicted in green, dark-green and light-green lines respectively, is shown for comparison. The three green shaded regions (green, dark-green and light-green) mark the DM mass ranges where the maximal transferred energy to nuclei is larger than to electrons, for H, He and $^6$Li, respectively. The violet shaded region mark the DM mass range where the maximal transferred energy to electrons is larger than the one to any nuclei, for the considered transferred momenta.}.
\end{figure}

The electronic energy levels in scintillators are in a superposition of states which distribute into two continuous bands separated by an energy gap, the valence and the conduction band. The electronic band structure and the electronic wave functions are calculated for each target using density functional theory (DFT) calculations. These functions can be obtained for instance by running the open source software QUANTUM ESPRESSO (QE)~\cite{QE} and used as inputs to open source codes such as QEdark~\cite{Essig:2015cda}, for the calculation of the DM-electron scattering rates.  DarkELF~\cite{Knapen:2021bwg} is an open source code that provides the DM-electron scattering rate including the dielectric function to consider in-medium screening effects~\cite{mermin1978solid, Gelmini:2020xir}. EXCEED-DM~\cite{Griffin:2021znd,PhysRevD.107.035035} is also an open source code independent from QE and can receive as input the electronic band structure calculated with any DFT software. It can compute the dielectric function and also include screening effects. Recently, an updated version of QEDark was released, called QCDark~\cite{Dreyer:2023ovnt}, which employes a different modelling of the electronic wave function calculation with respect to EXCEED-DM and provides similar results. The band structure of the ordered lattice of inorganic scintillators  and semiconductors which are solid state targets (\textit{e.g.} NaI, GaAs, Si, Ge) is different to the one in noble-gas liquids (\textit{e.g.} Xe, Ar), where the atoms are treated as isolated systems. 
In the case of electron inelastic scattering, the energy difference between the final and initial state of the scattered electron is estimated using the conservation of energy {\color{black} and the approximation that all energy is transferred to the electron, thus neglecting the atom recoil energy~\cite{Essig:2015cda,Derenzo:2016fse,Catena:2019gfa}},
\begin{equation}
\label{eq:encons}
\begin{cases}
E_i = \frac{1}{2}m_\chi {|\textbf{v}^{det}_{\chi,i}|}^2 + m_\chi + E_{e,i} \\
E_f = \frac{1}{2}m_\chi {|\textbf{v}^{det}_{\chi,f}|}^2 + m_\chi + E_{e,f}
\end{cases}
\end{equation}
 and it is equal to,
\begin{equation}
\label{eq:eenergy}
\Delta E_e = (E_{e,f}-E_{e,i}) = q v^{det}_{\chi,i}  \mbox{cos} \ \theta_{q v_{\chi,i}}- \frac{q^2}{2 m_\chi}
\end{equation}
where $q = |\textbf{p}-\textbf{p}'|$ is the transferred momentum, with $\textbf{p}$ ($\textbf{p}'$) the DM initial (final) momentum, {\color{black}{$\theta_{q v_{\chi,i}}$ is the angle between the transferred momentum and the incoming DM velocity vector in the detector reference frame, $v^{det}_{\chi,i}$ (indicated as $v^{det}_{\chi}$ in the following), and  $m_\chi$ is the DM mass. The maximal $\Delta E_e$, corresponding to $\mbox{cos} \ \theta_{q v_{\chi,i}} = 1$}} and calculated for a galactic escape velocity in the detector reference frame, $v^{det}_{\chi,esc} = 760$ km/s~\cite{Piffl:2013mla}, is shown in Fig.~\ref{fig:kinematics} as a function of the DM mass, for ten different transferred momenta, {\color{black} $q_i = i \alpha m_e$, with $i=1,...,10$,  where $v_e=i \alpha $ represents the speed of the electron and $i$ the effective nuclear charge, $Z_{eff}$~\cite{Essig:2015cda}}. In literature~\cite{Essig:2015cda,Derenzo:2016fse,Catena:2019gfa}, the transferred momentum is assumed equal to the electron momentum, because the electron velocity, $v_e\sim \mathcal{O}(\alpha c) \sim \unit[2000]{km/s}$, with $\alpha = 1/137$, is larger than the DM velocity, $v_\chi^{det} \sim 10^{-3} c \sim \unit[300]{km/s}$ and $m_e<m_\chi$. Thus, $q \simeq \mu_{\chi e} v^{det}_\chi \simeq m_e v_e$~\cite{Essig:2015cda}, where $\mu_{\chi e}$ is the DM-electron reduced mass. Using this approximation, in Fig.~\ref{fig:kinematics}, we compare the kinematics of DM-scattering off electrons with the one of DM elastic scattering off nuclei as a function of the DM mass. We show that the maximal-energy-transfer can be larger for electronic excitation than for H, He or $^6$Li nucleus recoils (depicted in solid green), for DM masses below about 90 MeV/c$^2$, 170 MeV/c$^2$ and 210 MeV/c$^2$, respectively. Helium and $^6$Li are the lightest target nuclei considered~\cite{SPICEHeRALD:2021jba} or employed~\cite{CRESST:2022jig} for DM search by the time of this work. It follows that DM-electron scatterings can correspond to larger energy depositions than DM-nuclei scatterings for $m_\chi \lesssim 90~\mbox{MeV/c}^2$.

The generic rate of electronic excitation from the valence to the conduction band in the crystal is~\cite{Catena:2021qsr},

\begin{equation}
\label{eq:rate}
    R_{crystal} = 2 \sum_{i i'} \int_{BZ} \frac{V d\textbf{k}}{(2 \pi)^3}\int_{BZ} \frac{V d\textbf{k}'}{(2 \pi)^3} R_{i \textbf{k}\rightarrow i' \textbf{k}'}
\end{equation}    

where the factor 2 comes from spin degeneracy, $i$ ($i'$) is the initial (final) electronic band index, $\textbf{k}$ ($\textbf{k}'$) the initial (final) wavevector in the first Brillouin zone (BZ), $V$ is the volume of the crystal. The rate of electronic transitions $R_{i \textbf{k}\rightarrow i' \textbf{k}'}$ is equal to~\cite{Catena:2021qsr},

\begin{equation}
\begin{aligned}
    R_{i \textbf{k}\rightarrow i' \textbf{k}'} = & \frac{n_\chi}{16 m^2_\chi m^2_e} \int \frac{d \textbf{q}}{(2 \pi)^3} \int d \textbf{v} \ f_\chi (\textbf{v}^{det}_\chi) \cdot \\ \cdot & (2 \pi) \ \delta(E_f - E_i) \overline{|M_{i \textbf{k}\rightarrow i' \textbf{k}'}|^2} 
\end{aligned}
\end{equation}

with $n_\chi=\rho_\chi/m_\chi$ the DM number density, $m_\chi$ ($m_e$) the DM (electron) mass, $f_\chi(\textbf{v}^{det}_\chi)$ the DM velocity distribution in the detector reference frame, and $\overline{|M_{i \textbf{k}\rightarrow i' \textbf{k}'}|^2}$ the scattering amplitude between the initial and final electron wave functions $\tilde{\psi}_{i \textbf{k}}$ and $\tilde{\psi}_{i'\textbf{k}'}$, integrated over the electron momenta, averaged over the initial spin states and summed over the final ones. In this notation, $\overline{|M_{i \textbf{k}\rightarrow i' \textbf{k}'}|^2}$ can include all possible DM-electron interactions~\cite{Catena:2021qsr}. If we select the first term in Eq.~(9) of Ref.~\cite{Catena:2021qsr}, 
$R_{crystal}$ in our Eq.~\ref{eq:rate} corresponds to $R_{crystal}$ in Eq.~(1) of Ref.~\cite{Derenzo:2016fse}. With this selection, the differential rate can be expressed as~\cite{Derenzo:2016fse},
\begin{equation}
\begin{aligned}
&\frac{d R_{crystal}}{d E_e} = \frac{\rho_\chi}{m_\chi} N_{cell}  \sigma_e  \alpha  \frac{m_e^2}{\mu^2_{\chi e}} \cdot \\ & \cdot
\int d  \mbox{ln} q  \left(\frac{1}{q} \ \eta(v_{min}(q, E_{e}))|F_{DM}(q)|^2 |f_{crystal} (q, E_e)|^2 \right)
\end{aligned}
\label{eq:rcryDerenzo}
\end{equation}

\begin{figure}
\includegraphics[width=\linewidth]{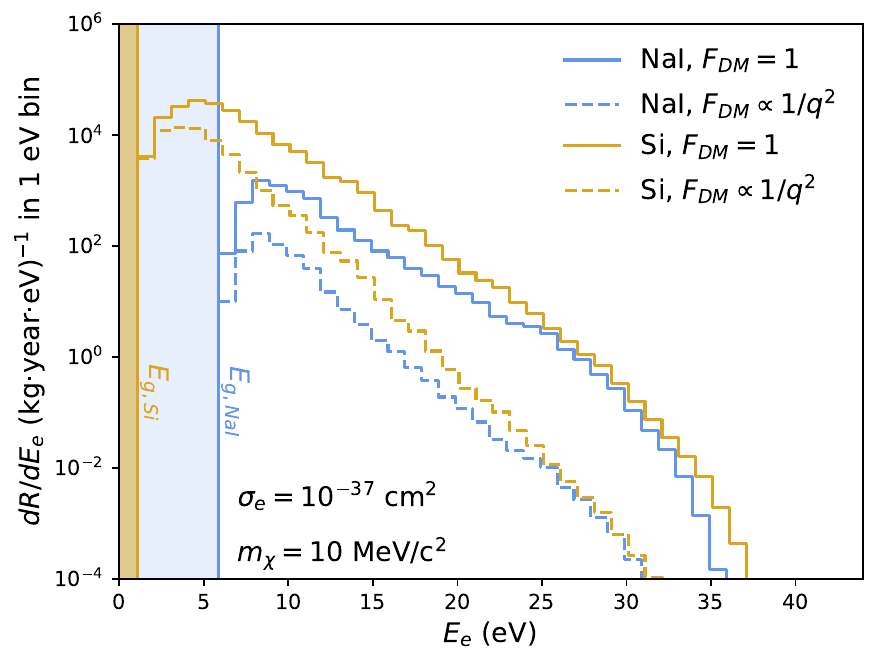}

\caption{\label{fig:DMe_rate} Differential DM-electron interaction rate per 1 eV bin and kg$\cdot$year in NaI (light blue) and Si (dark yellow), for $\sigma_e = \unit[10^{-37}]{cm}^2$, $m_\chi = \unit[10]{MeV/c}^2$ and for the energy gap $E_g = \unit[5.9]{eV}$~\cite{Derenzo:2016fse} (blue vertical line) and $\unit[1.1]{eV}$~\cite{Essig:2015cda} (dark yellow vertical line), respectively. The differential rate for $F_{\mbox{DM}} = 1$ is depicted with solid lines, while the one for $F_{\mbox{DM}} = (\alpha m_e/q)^2$ is depicted with dashed lines.}
\end{figure}

{\color{black} where $\rho_\chi$ is the DM mass density,} $N_{cell} = M_{target}/M_{cell}$ is the total number of unit cells, with $M_{cell} = N_{atoms/cell}M_{atomic}$ the mass of one unit cell, $\mu_{\chi e}$ is the DM-electron reduced mass, $\eta(v_{min}(q, E_{e}))$ is the velocity distribution integral, {\color{black} with  $v_{min}$ the minimum relative velocity required to induce an electron excitation of energy $E_e$}, $F_{DM}(q)$ contains information on the DM model, for instance it is $F_{DM}(q)\sim 1$ for heavy mediators ($m_{med} \gg \alpha m_e$) and $F_{DM}(q)\sim 1/q^2$ for ultralight mediators ($m_{med}\ll \alpha m_e$)~\cite{Essig:2015cda}, $f_{crystal} (q, E_e)$ is the crystal form factor and {\color{black}$\sigma_e$ is the cross-section of DM scattering off free electrons at fixed momentum transferred, } defined as,
\begin{equation}
\sigma_e = \frac{\mu_{e\chi}^2 \overline{|M_{free} (\alpha m_e)|^2}}{16\pi m_e^2 m_\chi^2}
\end{equation}

where $M_{free}(q)$ is the elastic scattering matrix element of DM with a free electron,  computed for $q=\alpha m_e$~\cite{Essig:2015cda,Derenzo:2016fse}. As an example, we show in Fig.~\ref{fig:DMe_rate} the differential rate of DM-electron interactions in NaI (in light blue) and Si (in dark yellow) calculated using a python version of QEdark\footnote{The Mathematica module of QEDark was re-written as a python notebook to employ all the crystals there provided.}~\cite{Essig:2015cda}, for (i) $m_\chi = \unit[10]{MeV}$, (ii) $F_{\mbox{DM}} = 1$  (solid line) and $F_{\mbox{DM}} = (\alpha m_e/q)^2$ (dashed line), \textit{i.e.} $\overline{|M_{i \textbf{k}\rightarrow i' \textbf{k}'}|^2}$ equal to the first term in Eq.~9 of Ref~\cite{Catena:2021qsr}, (iii) time of the year in the velocity distribution equal to March, (iv) $\sigma_e = \unit[10^{-37}]{cm}^2$ and (v) energy gap $E_g = \unit[5.9]{eV}$ and $\unit[1.1]{eV}$ in NaI and Si, respectively~\cite{Essig:2015cda}. The difference in the rate in NaI and Si comes from the different electronic configurations and band gaps. The DM model $(m_\chi, \sigma_e) = (\unit[10]{MeV/c^2}, \unit[10^{-37}]{cm^{2}})$ is used as a benchmark for comparison with previous studies\footnote{{Figure}~\ref{fig:DMe_rate} is in agreement with Fig.~4 of~\cite{Derenzo:2016fse}.}.


\section{Detection principle}

\subsection{Response of the scintillator}
\label{sec:response}
The mechanisms following the primary excitation of an electron into the conduction band induced by DM scattering in scintillators are equivalent to the ones following known scattering processes, thus widely studied. They can be divided in three sub-processes described e.g. in~\cite{https://doi.org/10.1002/pssb.2221870102},

\begin{enumerate}
\item (i) primary excitation, (ii) relaxation by creation of further electron (and hole) excitations, and (iii) thermalization of the produced electrons and holes into electron-hole-pairs of energy roughly equal to the energy gap, $E_g$;
\item (i) further relaxation by formation of excitonic states and (ii) energy migration to luminescence centers;
\item luminescence.
\end{enumerate}

The number of scintillation photons produced, $n_{\gamma}$, is,
\begin{equation}
\label{eq:ngamma}
n_{\gamma} = \frac{E_{e}}{E_{eh}} \ S \ Q
\end{equation}

where $E_{e}$ is the energy of the primary electron,  $E_{eh}$ is the energy necessary for the creation of an electron-hole pair at the band gap edge, $S$ is the efficiency of migration to luminescence centers and $Q$ is the yield of the final luminescent process. Note that $E_{eh} \neq E_g$, for $n_{\gamma} > 1$. The electrons lose energy in the thermalization into electron-hole-pair formation, a material-dependent mechanism described by the energy loss function~\cite{https://doi.org/10.1002/pssb.2221870102, Knapen:2021bwg}.\\

To the aim of employing the scintillator as a target for DM-electron interactions, the scintillator response to an electron recoil must be studied. Exemplary, the measurement of the energy loss function for the specific crystal, ideally at low energies (below $\unit[50]{eV}$) and at the operating temperature ($\sim \unit[10]{mK}$), is necessary as the threshold energies for the production of a specific number of scintillation photons, $n_\gamma$, depend on it. Similarly, experimental studies, ideally at the energy and temperature of interest, of the parameters $S$ and $Q$ are required to perform an actual energy and detector efficiency reconstruction. In the following we fix {\color{black} $E_{eh} = E_g + (n_{\gamma}-1)\langle E \rangle$,  where $\langle E \rangle$ is the mean average energy required to produce another electron-hole pair, here fixed to $3E_g$, according to~\cite{Essig_2012}}. 
Under this assumption, the list of threshold energies $E^{th}_{eh}$ is $(E_g, 4E_g, 7E_g, 10 E_g, \dots)$ for the production of $n_\gamma = (1,2,3,4, \dots)$ scintillation photons, respectively.

\subsection{Observables}
CSCs employed for rare event searches~\cite{COSINUS:2023kqd,CRESST:2022jig,NUCLEUS:2019igx} are realized by combining a scintillating target crystal with good phonon propagation properties (e.g. with high Debye temperature, low impurity concentration and lattice defects, polished surfaces) with a second crystal absorbing the scintillation light of the target. The target crystal is coupled to a sensitive thermometer to measure the phonon signal generated by the primary interaction in the target (phonon detector {\color{black} channel}). The scintillation light absorbed by the second crystal is measured using a thermometer placed on its surface; the absorber of the scintillation and its sensor constitute the cryogenic light detector {\color{black} (or light channel)}. The double channel readout of phonon and light is a successful technique to perform particle identification and electromagnetic background rejection on an event-by-event basis in the search for DM-nucleus interactions~\cite{CRESST:2022jig,COSINUS:2023kqd}. The energy measured in the phonon channel $(E_p)$ gives a measure of the total deposited energy in the target. The coincident energy measured in the light channel $(E_l)$ depends on the type of interaction (electromagnetic vs elastic nuclear recoil). The ratio of the two energies, $E_l/E_p$, enables event-by-event particle discrimination.\\

The ensemble of scintillating absorber and light detector is extremely suitable for investigating the presence of DM-electron interactions in DM direct search experiments. Indeed, scintillators are characterized by electronic band gaps of ~$\mathcal{O}(\mbox{eV})$. The primary DM-electron interaction described in Sec.~\ref{sec:DMe} is expected to trigger the scintillation process, thus the emission of photons of energy equal to the band gap size, $E_g$~\cite{Derenzo:2016fse}. The observable would be single photon peaks, detectable with a cryogenic light detector with eV/sub-eV energy resolution.

In ref.~\cite{Derenzo:2016fse}, the light detector is an array of sensors placed on the target surface itself, a suitable choice to reach the required $\mathcal{O}$(eV) energy threshold. {\color{black} However, in such layout the sensors would be sensitive to any phonons propagating from the absorber to the sensor. The backgrounds originating for instance by external sources scale with the target mass, and consequently, with the exposure to the signal. Additionally, the hypothesis that the background at low energy scales with the size of the sensor to absorber interface due to thermal stress, is under discussion, see e.g.~\cite{fuchs23}. Furthermore, the baseline resolution of the cryogenic detectors is found to worsen when increasing the absorber size~\cite{Strauss:2017cam}.} By separating the light detector from the scintillator, the background can be mitigated by decreasing the mass of the light detector without affecting the exposure {\color{black} or the performance}. The configuration of the CSCs optimized for the DM-nucleus scattering search fulfills this requirement, but the required sensitivity of the cryogenic light detectors was not yet demonstrated when this idea was first proposed.\\

Recently, the sensitivity to DM-electron interactions has been shown to be in reach of cryogenic light detectors read out with transition-edge-sensors: the first single photon detection in a CRESST light detector was, indeed, proven~\cite{CRESST:2024cpr}. Inspired by Refs.~\cite{Derenzo:2016fse} and~\cite{CRESST:2024cpr}, here we propose to optimize the cryogenic light detectors which are already operated in several cryogenic calorimeters~\cite{COSINUS:2023kqd,CRESST:2022jig,NUCLEUS:2019igx} and employ them for the detection of single photons emitted after primary DM-electron interactions in scintillating target crystals.\\

The expected energy in the light ($E_{l}$) and phonon ($E_{p}$) detectors, under the assumption that the energy not converted into scintillation is converted into phonons, are,

\begin{equation}
\begin{cases}
E_{l} = n_{\gamma} \ E_g \\
E_{p} = E_e - E_{l}
\end{cases}
\end{equation}

We recall that $n_\gamma$ is defined in Eq.~\ref{eq:ngamma}, where $E_{eh}$ is the threshold energy for the creation of an electron-hole pair at the band gap edge,  $E_g$ is the electronic energy gap, $E_e$ is the energy of the primary electron and the relation among these that we consider is $E_{eh} = E_g$ for $n_{\gamma} = 1$, and $E_{eh} = E_g + (n_{\gamma}-1)\expval{E}$, with $\expval{E} =3E_{g}$, for $n_{\gamma} > 1$, according to~\cite{Essig_2012}. Following the assumptions on the scintillator response given in Sec.~\ref{sec:response}, the number of emitted photons, $n_\gamma$, which corresponds to the number of created electron-hole pairs, is,

\begin{equation}
n_{\gamma} = 1 +\lfloor{(E_e-E_g)/\expval{E}}\rfloor, \mbox{ for } E_e \geq E_g
\end{equation}

\begin{figure}[htb!]
\includegraphics[width=\linewidth]{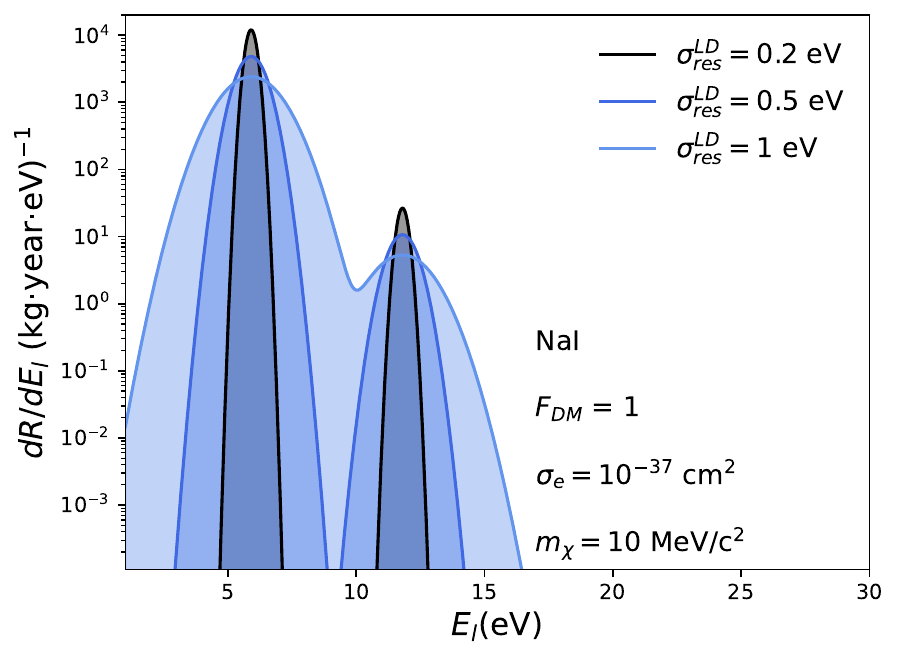}
\caption{\label{fig:peaks} Rate of scintillation photon absorbed as a function of the deposited energy in the light detector, for DM-electron interactions in NaI, under the assumption of the DM scenario shown in Fig.~\ref{fig:DMe_rate} and $F_{DM}=1$, for $E_{g,NaI} = \unit[5.9]{eV}$, $\expval{E}=3E_g$ and assuming three different energy resolutions of the light detector (LD), $\sigma_{res}^{LD}=\unit[0.2]{eV}$ (black), $\sigma_{res}^{LD}=\unit[0.5]{eV}$ (blue) and $\sigma_{res}^{LD}=\unit[1]{eV}$ (light-blue).}
\end{figure}

 The signature of DM-electron interactions in a light detector with gaussian energy resolution, $\sigma_{res}$, is peaks centered at multiple energies of $E_g$, as shown in Fig.~\ref{fig:peaks}, for $\sigma_{res} = 0.2, 0.5$ and $1$~eV (black, blue and light-blue, respectively). This plot is obtained using the same DM model as in Fig.~\ref{fig:DMe_rate}, the energy gap $E_{g,NaI} = \unit[5.9]{eV}$, the ideal assumption of {\color{black} scintillation light detection efficiency\footnote{\color{black} We label \textit{scintillation light detection efficiency} the light emission efficiency times the light collection efficiency.}} equal to 1 and zero background (the role of the background is discussed later). The performance scenarios for the energy resolution are based on the state-of-the-art cryogenic light detectors. Exemplary, CRESST-III has achieved the high baseline resolution of \unit[1.0]{eV} in a silicon-on-sapphire (SOS) light detector~\cite{CRESST:2024cpr}. 
We assume that the total energy not converted into scintillation converts entirely into phonons. 

\subsection{DM mass estimation}

\begin{figure}
\includegraphics[width=\linewidth]{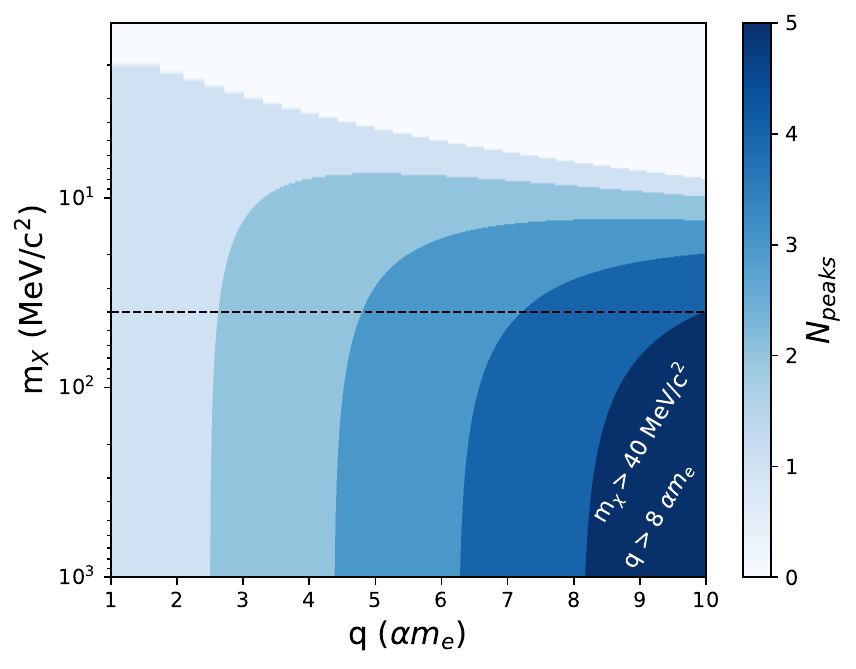}
\caption{\label{fig:DMmassident} 2D color plot of the number of peaks expected in the light detector (color map) as a function of the DM mass (left y-axis) and transferred momentum (x-axis), in a NaI crystal with $E_g = \unit[5.9]{eV}$ and for $v_{\chi,esc}^{det} = \unit[760]{km/s}$. The dark blue region corresponds to five peaks in the light detector and to a DM mass larger than 40 MeV/c$^2$, as marked by the black dashed horizontal line.}
\end{figure}

\begin{figure}
\includegraphics[width=\linewidth]{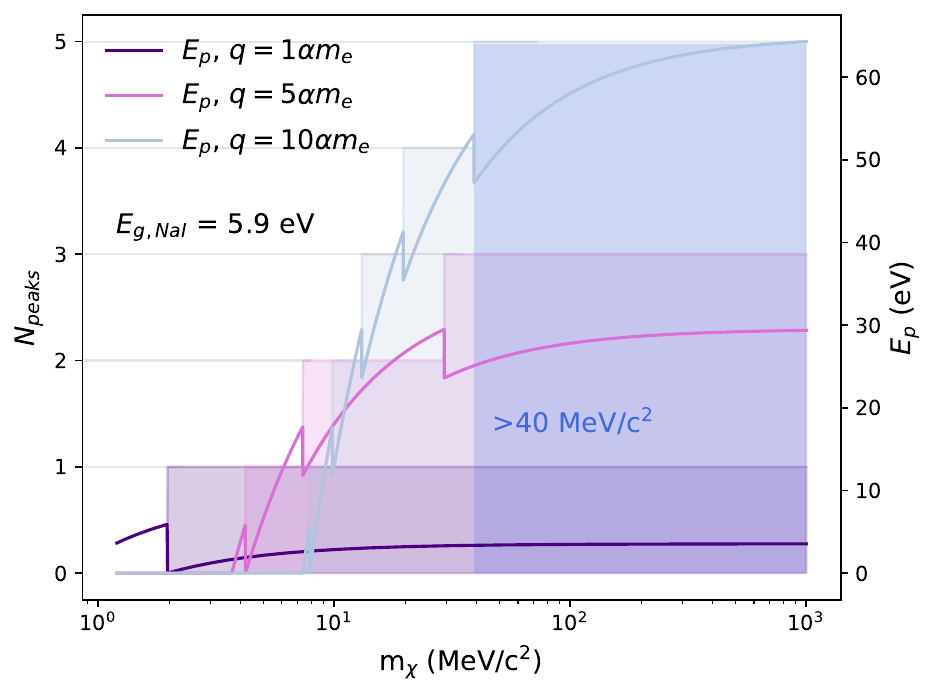}
\caption{\label{fig:DMmassident_phononchannel} Number of observed peaks in the light detector (left y-axis) and corresponding energy in the phonon channel (right y-axis), as a function of the DM mass. The three shaded regions (violet, pink and light blue) show the peaks in the light detector for three transferred momenta $q_i= i\alpha m_e$, with $i = 1,5$ and $10$, respectively, as a function of the DM mass. For a fixed DM mass, the highest step of each shaded region corresponds to the number of peaks in the light detector for $q_1, q_5$ and $q_{10}$, respectively. The violet, pink and light blue solid lines depict the energy going into phonons, for the three transferred momenta $q_1, q_5$ and $q_{10}$, respectively, as a function of the DM mass. The blue shaded region marks the DM mass region ($m_\chi > \unit[40]{MeV/c}^2$) which would be compatible with an observation of five peaks in the light detector. The corresponding energy in the phonon channel would be a value along the solid light-blue line in the blue-shaded region, between 47.6 eV for $m_\chi = \unit[40]{MeV/c}^2$ and 64.3 eV for $m_\chi = \unit[1000]{MeV/c}^2$.}
\end{figure}

When fixing the detector response parameters $S$ and $Q$ in Eq.~\ref{eq:ngamma}, the number of emitted scintillation photons depends on the transferred momentum, $q$, and the DM mass, $m_\chi$.  For a specific DM mass, the spectrum of energy of the recoiling electron is a series of peaks seen in the light channel. 
Considering the uncertainty on the primary electron momentum, a quantity which allows to estimate the range of DM masses compatible with the signal is the number of observed peaks in the light detector. This method is shown in Fig.~\ref{fig:DMmassident}, where the number of peaks is depicted using the color map on the right y-axis and estimated as a function of the DM mass and transferred momentum. For instance, if five peaks were observed, using Fig.~\ref{fig:DMmassident} it would be possible to conclude that the DM particle mass is larger than about 40 MeV/c$^2$ (dark blue region) and transferred momenta above $q=8\alpha m_e$ are feasible. 
We find that an extremely sensitive phonon detector with {\color{black} sub-eV energy resolution} {\color{black} would be needed} to improve the estimation of the DM mass. This is shown in Fig.~\ref{fig:DMmassident_phononchannel}. We plot as a function of the DM mass the number of peaks in the light channel on the left y-axis  (the steps depicted as shaded regions) and the maximal energy in the phonon channel on the right y-axis (the solid lines, calculated for $v_{\chi,esc}^{det} = \unit[760]{km/s}$). Using the previous example, if five peaks were observed, the maximal energy measured in the phonon channel in coincidence to the events in the fifth peak of the light channel, would correspond to a specific DM mass, as visible in Fig.~\ref{fig:DMmassident_phononchannel}. 

\subsection{Envisioned Detector Design}

Modifications to the CSC design state-of-the-art must be conceived to optimize the detectors for the DM-electron interaction search. To maximize the scintillation collection efficiency, a full coverage of the target crystal is required. A key rule of calorimetry is that the sensitivity degrades while increasing the heat capacity, thus the calorimeter mass, see e.g.~\cite{Strauss:2017cam}. CRESST-III achieved a baseline resolution of 1~eV with a $(20\times20\times0.4)$~mm$^3$ silicon-on-sapphire wafer~\cite{CRESST:2024cpr}. An array of wafers all equipped with transition-edge-sensors which can be read out as a single channel can provide both the $4\pi$-coverage and the baseline resolution necessary to detect the single photon peaks. 
Regarding the required performance of the phonon channel, we already showed that a phonon channel with sub-eV baseline resolution {\color{black} (thus of small target mass) would be needed to pin down the DM mass, but a range of values can be identified using the number of peaks detected in the light channel. This implies that the readout of the phonon channel is not required for signal detection but only for model selection. The detection in coincidence of light and phonon signal can discriminate background events, such as direct hits in the light detector, from the signal, but not below the energy threshold of the phonon channel. In order to have a phonon detector with comparable performance to the light detector, again, a small target mass would be necessary. From these observations it follows that, to the purpose of signal detection,} the sensitivity of the phonon channel can be traded with a larger exposure to the signal, that is with an increase of the target mass. In Sec.~\ref{sec:projections}, {\color{black} the sensitivity projections for 1 kg$\cdot$year of exposure and different background levels and detection efficiencies will be estimated}.

\subsection{Method for Energy Gap Measurement at mK }
\begin{figure}[htb!]
\includegraphics[width=\linewidth]{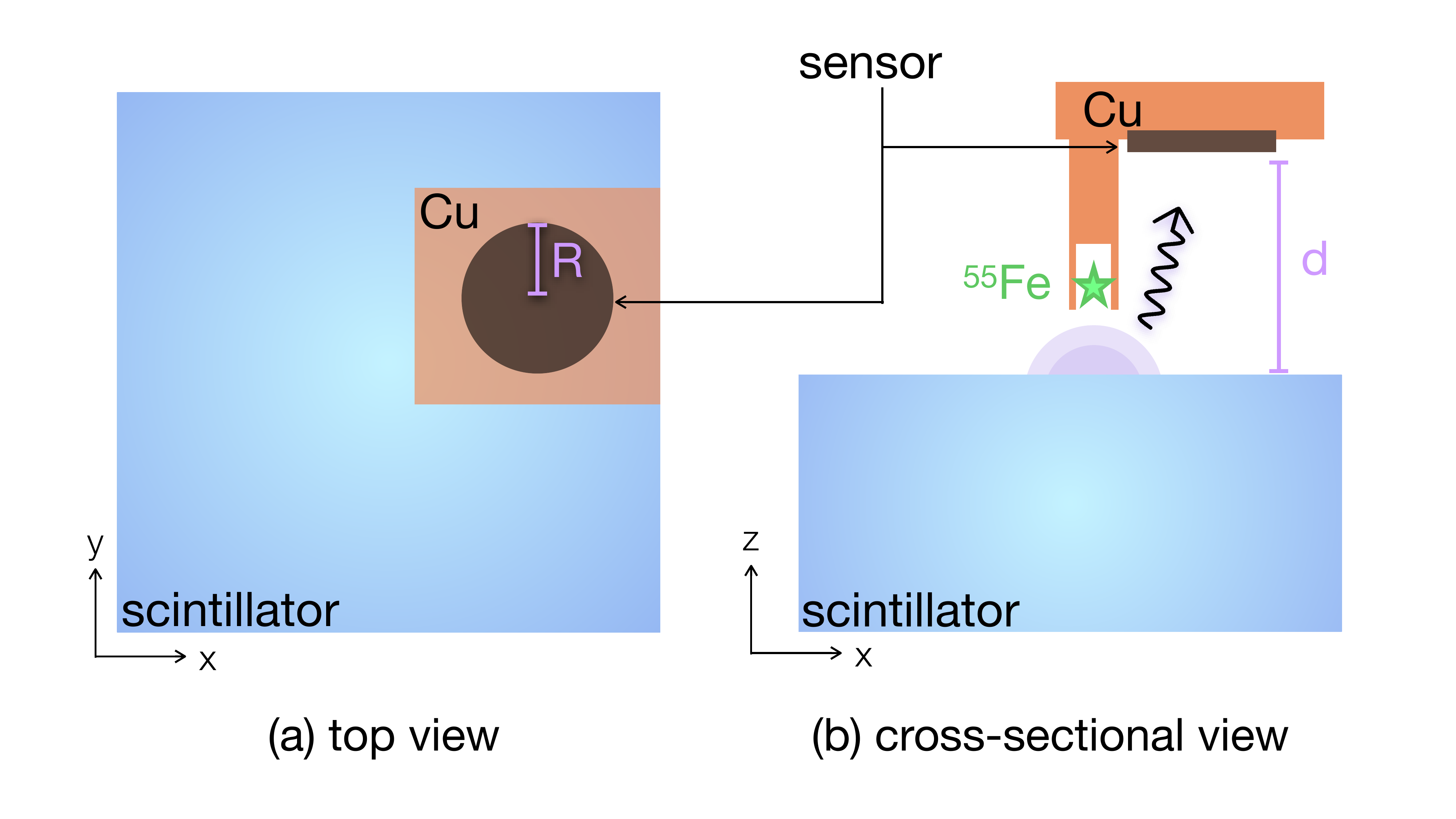}
\caption{Schematic of the setup for the measurement of the energy gap at mK. {\color{black} In the top view (a) the copper holder is depicted transparent to show the sensor, {\color{black} illustrated as a brown disk}. In the cross-sectional view (b) the {\color{black} collimated} iron source is depicted as shielded from one side, inserted into a screw. The violet circles represent the scintillation emission.}\label{fig:egapmeasurement} }
\end{figure}

The energy gap between the conduction and valence electronic bands is a characteristic of the scintillator which depends on the atomic composition and other properties, like the presence of impurities or dopants. For the same target material, different energy gap values can be found in literature~\cite{Derenzo:2016fse, rotman1997wide,moszynski2002study,J_B_West_1970}. Thus, a dedicated measurement of the energy gap of the target employed is crucial for this research. To this aim, we propose to use as {\color{black} light sensor} a superconducting foil (or a normal-superconducting bilayer) connected to a sensor, for example using the remoTES design~\cite{COSINUS:2021onk} or directly a thermometer, e.g. a TES~\cite{jodoi2022iridium}. The requirements for this measurement are a calibrated sensor with sensitivity of $\mathcal{O}$(eV) to direct hits and a target with known light yield. We will consider $\epsilon_l = 13\%$ of the total deposited energy in the target for pure NaI crystals, which is the percentage of energy in light detected by the beaker-shaped silicon light detector encapsulating a pure NaI crystal in Ref.~\cite{COSINUS:2017bco}. A scheme of the setup is given in Fig.~\ref{fig:egapmeasurement}.

        

A {\color{black} light sensor} is fixed to a copper holder. An X-ray source, e.g.~$^{55}$Fe which emits 5.9~keV and 6.4~keV from the K-$\upalpha$ and K-$\upbeta$ atomic shells, is placed below the other face of the scintillator. A detection of $\epsilon_l = 13\%$ of the deposited energy in form of scintillation light, corresponds to about 780 eV for an average X-ray energy of $E_{cal} = 6$~keV. The goal is to lower the total collection efficiency in the temporary light detector to increase the probability of the single photon detection. In this way, the 1$\gamma$-peak is intensified and the estimation of the mean of the peak is directly the estimation of the energy gap. If multiples of these peaks are also observed, the energy gap estimation can be further refined using the distance between consecutive peaks~\cite{CRESST:2024cpr}. For instance, in the configuration of Fig.~\ref{fig:peaks}, a detected energy $E_{det} = 8$~eV would be a good choice, since it is larger than $E_g= 5.9$~eV and lower than $2E_g= 11.8$~eV. In this way, the temporary light detector is reached by a high rate of single scintillating photons and the mean of the resulting peak is the scintillator energy gap. The single photon detection with a Au-Ir superconducting bilayer was recently demonstrated with a monocromatic source at \unit[850]{ nm}~\cite{jodoi2022iridium}. We propose a similar experiment, tuned to the lower wavelength window of (200-400)~nm where the pure NaI scintillation at low temperature is expected, as part of the proof-of-principle of the calibration method. The geometry of the setup in Fig.~\ref{fig:egapmeasurement} can be adapted to this purpose by using the following formula,

\begin{equation}
\label{eq:egapmeasurement}
E_g < E_{cal} \cdot \epsilon_l \cdot \frac{A_{sensor}}{4 \pi d^2} \leq E_{det}
\end{equation}

where $A_{sensor}$ is the surface area of the sensor and $d$ is the distance of the sensor from the scintillator surface. For a superconducting foil of radius $R$ as in Fig.~\ref{fig:egapmeasurement}, Eq.~\ref{eq:egapmeasurement} gives the following relation between the size of the superconducting foil and the required distance from the scintillator,

\begin{equation}
\label{eq:dvsR}
\sqrt{\frac{E_{cal} \cdot \epsilon_l}{E_{det}}} \frac{R}{2} \leq d < \sqrt{\frac{E_{cal} \cdot \epsilon_l}{E_{g}}} \frac{R}{2}
\end{equation}

{\color{black} In case of inhomogeneous wavelength of the emitted scintillation light, the measurement could suffer from a worsened resolution of the peaks. If the broadening is lower than the distance between two consecutive peaks, the peaks can still be distinguished.}

\subsection{Expected Background}
\label{sec:bck}
The background in searches for DM-electron interactions using CSCs originates both from the scintillating target and from the light detector. Starting from the former, the energy deposited by ionizing radiation, charged and neutral particles in the target, can produce scintillation light, which would distribute in the same peaks as the ones populated by the DM events. All the surroundings (light detector included) can emit particles, mainly photons, interacting in the target and leading to scintillation. The dominant atomic interaction with photons at energies below about 1 keV is the photoelectric absorption~\cite{hubbell1980pair, Workman:2022ynf}. A dedicated simulation of the surroundings is necessary to model the scintillation background and take it into account in a likelihood framework. \\

\begin{figure}[htb!]
\includegraphics[width=\linewidth]{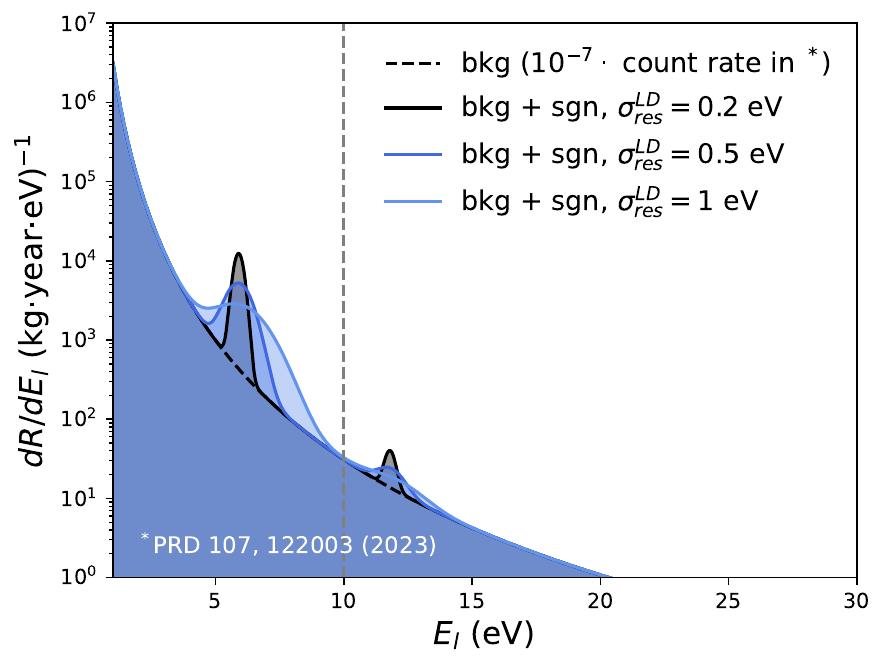}
\caption{\label{fig:sgnbck} The DM-electron interaction peaks detected in the light detector, for a detector resolution equal to $\sigma_{res} = \unit[0.2]{eV}$ (black), $\sigma_{res} = \unit[0.5]{eV}$ (blue) and  $\sigma_{res} = \unit[1]{eV}$ (light-blue). The background rate is taken from~\cite{PhysRevD.107.122003}, Fig.~4, suppressed by a factor $10^{-7}$ for visualization. The dashed vertical line signs the lowest data point energy in ~\cite{PhysRevD.107.122003}, below the curve is the extrapolation of the fit function. The signal is calculated according to the same assumptions as in Fig.~\ref{fig:DMe_rate}.}
\end{figure}

The second type of background originates from any type of direct energy deposition in the light detector, since the absorbed light is converted into phonons and then measured by the {\color{black} phonon-mediated} sensor. The expected background in the light detector currently represents a serious limitation to this experiment. Cryogenic experiments have reached very low energy thresholds $\mathcal{O}$(eV). In these energy regions cryogenic calorimeters observe {\color{black} a low-energy excess (LEE) of events}, which does not have a clear explanation yet~\cite{Fuss:2022fxe}. In Fig.~\ref{fig:sgnbck} we show the photon peaks induced by DM particles of \unit[10]{MeV/c$^2$} with $\sigma_e = 10^{-37}$~cm$^2$, as assumed for previous plots, and we add the LEE detected by CRESST-III {\color{black} with a silicon wafer of 10 eV energy threshold ~\cite{PhysRevD.107.122003}, suppressed by a factor $10^{-7}$ and extrapolated below 10 eV}.  A milder suppression of the current background of CRESST-III is required to explore a new parameter space, as discussed in Sec.~\ref{sec:projections}, but in Fig.~\ref{fig:peaks} it is emphasized for visualization purposes. An exceptional suppression of the LEE is expected to be required for accessing a new DM-electron parameter space. {\color{black} However, considering light DM particles ($<\unit[100]{MeV}/$c$^2$), the impact of the LEE on the CSC sensitivity is less severe for DM-electron scattering search than for the DM-nuclei scattering search, since for the same light DM mass the expected number of events above threshold is larger for the former. Additionally, the experimental configuration proposed in this work has a two-fold advantage: (1) a peak search over a monotonically decaying background is expected to improve the signal to noise discrimination in a likelihood framework; (2) the target mass, thus the exposure, can be scaled up freely if the phonon channel is not considered with no reduction of the energy baseline resolution, a novelty for DM search using cryogenic calorimeters. However, a larger target mass can affect the light emission efficiency due to self-absorption. }{\color{black} This effect is an inherent property of scintillators and must be measured to be quantified. We have not yet investigated it.}

\section{SIGNAL DISCOVERY POTENTIAL}
\label{sec:projections}
\subsection{Statistical method}

The potential for signal discovery is studied using a profile-likelihood (PL)-based test and Monte Carlo (MC) data~\cite{Cowan:2010js}. The null-hypothesis $\mathcal{H}_0$ is the background-only model, the alternative-hypothesis $\mathcal{H}_a$ is the signal-plus-background model. Data are labelled as $\textbf{d}_{\mathcal{H}_0}$ and $\textbf{d}_{\mathcal{H}_a}$, if simulated under the null- or the alternative-hypothesis, respectively. The probability density function of signal-plus-background is,

\begin{equation}
\label{eq:sgnplusbck}
f(E_{l}, \sigma_e, \overline{m}_\chi, \Theta) = \frac{\sum_n f_{pk,n}(E_{l}, \sigma_e, \overline{m}_\chi) + f_{bkg}(E_{l}, \Theta)}{N_{tot}}
\end{equation}
where $\sigma_e$ is the strength parameter, thus the parameter of interest, $m_\chi$ is fixed ($\overline{m}_\chi$), {\color{black} $\Theta$ indicates the nuisance parameters corresponding to the background model $f_{bkg}(E_{l}, \Theta)$, and  $f_{pk,n}(E_{l}, \sigma_e)$ and $N_{tot}$ are,}

\begin{multline}
    f_{pk,n}(E_{l}, \sigma_e, \overline{m}_\chi) = \int dE' R_{n\gamma}(\sigma_e,\overline{m}_\chi)\cdot \\ \cdot \delta(E'-nE_{g})\  \phi (E_l - E')
\end{multline} 

\begin{equation}
N_{tot} = \int_\Delta dE_l \ \sum_n f_{pk,n}(E_{l}, \sigma_e, \overline{m}_\chi) + f_{bkg}(E_{l},\Theta)
\label{eq:ntot}    
\end{equation}

where {\color{black} $R_{n\gamma}$ is the number of events in the $n$-peak, $\phi(E_l-E')$ a gaussian distributed energy resolution with the width equal to $\sigma_{res}^{LD}$}, 
and $\Delta$ is the energy range considered. The test statistics used is,

\begin{align}
q = 
\begin{cases}
- 2\ \mbox{ln}\left(\frac{\mathcal{L}(\textbf{d}_{\mathcal{H}}|\sigma_e =0, \hat{\hat{\Theta}})}{\mathcal{L}(\textbf{d}_{\mathcal{H}}|\hat{\sigma}_e, \hat{\Theta})} \right), \quad & \hat{\sigma}_e \geq 0 \\
0, \quad &\hat{\sigma}_e < 0
\end{cases}
\end{align}
where $q \equiv q_0$ for $\mathcal{H} \equiv \mathcal{H}_0$ and $q \equiv q_a$ for $\mathcal{H} \equiv \mathcal{H}_a$ and $\mathcal{L}$ is the binned likelihood,

\begin{equation}
\label{eq:likelihood}
\mathcal{L}(\textbf{d}_{\mathcal{H}}|\sigma_e, \Theta) = \prod_i \frac{e^{\nu_i}}{n_i!} \nu_i^{n_i}
\end{equation}

where $\nu_i = N_{tot}\int_{E_{l,i}}^{E_{l,i+1}} \mbox{d}E_l\ f(E_{l}, \sigma_e, \overline{m}_\chi, \Theta)$ {\color{black} is the expected number of events} and $n_i$ is the number of simulated events falling in the $i$-bin. 
For a fixed DM-mass, several data-sets can be simulated under $\mathcal{H}_0$ and the $q_0$ can be calculated for each of them, thus obtaining a distribution of $q_0$, $f(q|\mathcal{H}_0)$. However, since $H_0$ is the background-only hypothesis, $f(q|\mathcal{H}_0)$ can be well approximated by a half chi-squared distribution, $1/2 \chi^2$~\cite{Cowan:2010js}, with one degree of freedom. We verified that the $1/2 \chi^2$ distribution fits well the simulated distribution of $q_0$ for a combination of DM mass and cross-section, and used the analytical equation of a $1/2 \chi^2$ for the $f(q|\mathcal{H}_0)$. With this approximation, we save the computational cost of the $f(q|\mathcal{H}_0)$ for each cross-section and mass. We then simulated 1000 data-sets under $\mathcal{H}_a$.  $\mathcal{H}_a$ is fixed by selecting a DM mass in the range $[\unit[2]{MeV/c}^2-\unit[1]{GeV/c}^2]$ and a cross-section, $\sigma_e$. This latter is treated as a free parameter in the minimization of the negative log-likelihoods in Eq.~\ref{eq:likelihood}. For each of the 1000 realizations, we compute the $q_a$.  The median, $q_{med}$, of the distribution $f(q|\mathcal{H}_a)$ is compared to $q_\alpha$, where $q_\alpha$ is fixed by the equation,

\begin{equation}
\label{eq:qalignement}
1-\alpha = \int_{q_\alpha}^{\infty}  \mbox{d}q \ f(q|\mathcal{H}_0) 
\end{equation}

where $\alpha\cdot 100\%$ is the desired confidence level (CL). If $q_\alpha \equiv q_{med}$, we can reject the null-hypothesis with $\alpha\cdot 100\%$ CL. 

\subsection{Sensitivity projection}

\begin{figure*}
\includegraphics[width=0.48\linewidth]{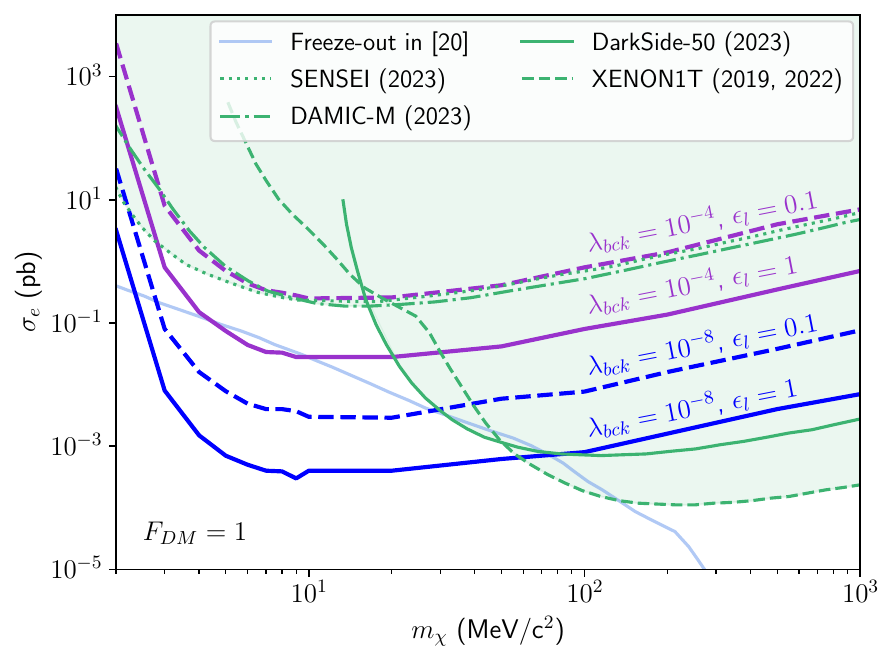}
\includegraphics[width=0.48\linewidth]{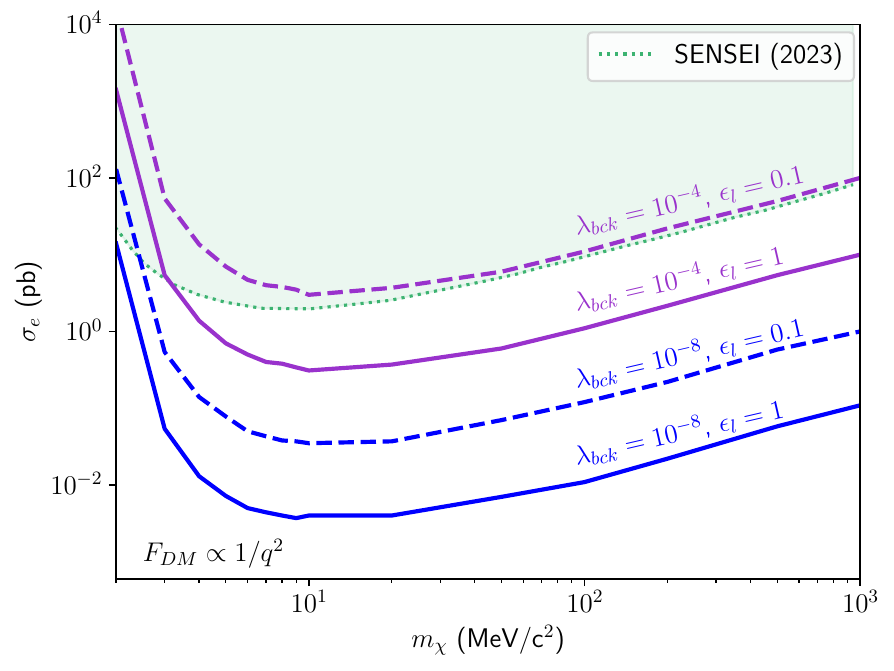}
\caption{\label{fig:sensitivity} Projection of the sensitivity to DM-electron interactions in a dedicated NaI cryogenic scintillating calorimeter, for $F_{DM}=1$ (left) and $F_{DM} \propto 1/q^2$ (right), $E_{g,NaI} = \unit[5.9]{eV}$ and $\expval{E} = 3E_{g,NaI}$, calculated for 1 kg $\cdot$ year exposure. {\color{black} The background rate is taken from~\cite{PhysRevD.107.122003} and suppressed by a factor  $\lambda_{bck} = 10^{-4}$ (purple curves) and $\lambda_{bck} = 10^{-8}$ (blue curves). The projections are calculated for a {\color{black} scintillation light detection efficiency} of 0.1 (dashed curves) and 1 (solid curves). The current experimental sensitivity, depicted as a shaded green area, is extracted from~\cite{DAMIC-M:2023gxo, SENSEI:2023zdf} and includes results from SENSEI~\cite{SENSEI:2023zdf}, DAMIC-M~\cite{DAMIC-M:2023gxo}, DarkSide-50~\cite{DarkSide:2022knj} and XENON1T~\cite{aprile2019light,XENON:2021qze}. For the light mediator scenario, Ref.~\cite{Bhattiprolu:2023akk} reports about a mismatch of their results on the freeze-in curve with the results of~\cite{Essig_2012}, which is the one shown in~\cite{DAMIC-M:2023gxo}.}}
\end{figure*}

The comparison of $q_\alpha$ and $q_{med}$ described in the previous section is made by scanning on the cross-section, $\sigma_e$, in different ranges for the different DM masses, from lower to larger cross-sections. We fix our desired CL to 90\%, corresponding to $q_\alpha\simeq1.63$ ($\int_{q_\alpha}^{\infty}  \mbox{d}q \ f(q|\mathcal{H}_0) \simeq 0.10$) and check the condition  $q_{med} > 1.63$. We label the cross-section of the simulation which satisfies this condition ``$\sigma^*_i$", where $i$ is the scanning index. We compute a finer scanning between $\sigma_{i-1}$ and $\sigma^*_{i}$, using 1000 simulations for each intermediate cross-section and define the discovery limit at the cross-section in this interval which corresponds to the condition {\color{black} $1.63 <q_{med} <2$}, namely to a conservative confidence interval larger than 90\% {\color{black} ($q_{med}\simeq1.63$)}, but lower than 95\% {\color{black}($q_{med}\simeq2.71$)}. With this method a limit at 90\% CL is statistically conservative at the advantage of the computational time. The background is simulated using $f_{bkg}(E_{l}, \Theta)$ {\color{black} equal to,

\begin{equation}
    f_{bkg}(E_{l}, \Theta) = \lambda_{bck} \cdot (A \  E_l^{-\alpha} + B \  E_l^{-\beta})
    \label{eq:fbck}
\end{equation}

 where $\lambda_{bck}$ is} a suppression factor. We assume that the shape of the background is known, thus we fix the parameters $A, \alpha, B, \beta$ to the estimated values reported in TABLE I of~\cite{PhysRevD.107.122003} {\color{black} and we leave the scaling factor $\lambda_{bck}$ as a nuisance parameter.
The results shown in Fig.~\ref{fig:sensitivity} are obtained for $\lambda_{bck} = 10^{-4}$ and $10^{-8}$ and {\color{black} scintillation light detection efficiency}, $\epsilon_l$, equal to 10\% and 100\%. A mitigation of a factor about $\lambda_{bck} = 10^{-4}$ is required to probe new parameter space. In the current status of development, the achievement of such background suppression is optimistic}. Different groups are focusing their efforts to study and reject this background. The level recorded by CRESST~\cite{CRESST:2022jig} and used here as reference is considered likely to be of the same origin as the ones of other cryogenic experiments and a common effort to reduce it was established~\cite{Fuss:2022fxe}. 

\section{CONCLUSIONS}
The CSCs have been employed so far for the search of DM interacting with nuclei. This work sets the ground for the employment of the CSCs in the search of DM particles interacting with electrons. We show that the phonon-channel is not required for the {\color{black} detection} of the signal, which instead can be {\color{black} observed} in the light channel in the form of peaks on top of the expected background. We discuss a method for the measurement of the electronic band energy gap which does not require the installation of an optical window in the cryostat. We also discuss the role of the low energy excess (LEE) observed in all cryogenic detectors on this search and plot {\color{black} projected sensitivity curves for a LEE reduction by a factor $10^{-4}$ and $10^{-8}$ with respect to the LEE recorded in a CRESST-III silicon wafer detector}, which show the potential contribution of the CSCs to the search of DM-electron interactions. This work is a reference for the performance of CSCs required to open a new frontier of research with a very established technology. Developments of these ideas are already ongoing and will be topic of future publications.

\begin{acknowledgments}

We are very grateful to Riccardo Catena, Dominik Fuchs, Maximilian Gapp, Gonzalo Herrera, Federica Petricca and Florian Reindl for the interesting discussions related to this study. We additionally thank Riccardo Catena, Dominik Fuchs and Gonzalo Herrera for their comments on a draft of this article, which enriched the final manuscript. {\color{black} We are thankful to the Klaus Tschira Foundation for supporting this research.}

\end{acknowledgments}


\bibliography{refs.bib}
\end{document}